\documentstyle[12pt]{article}
\thicklines
\begin{document}
\begin{center}
{\Large\bf On exact solution for some integrable nonlinear equations of
the Schr\"odinger type.}\\[0.5cm]
{\large V.G. Makhankov\\[0.3cm]}
{\em  Center for Nonlinear Studies, Los Alamos National Laboratory, P.O.Box
1663, Los Alamos, N.M. 87545, USA  \\
and Joint Institute for Nuclear Research, LCTA, Head Post Office, P.O. Box 79,
Dubna, Moscow 101000, Russia} \\
\vspace {0.8 cm}
\end{center}
\begin {abstract}
\item The outlook of a simple method to generate localized (soliton-like)
potentials of  time-dependent Schrodinger type equations is given. The
conditions are discussed for the potentials to be real and nonsingular.
For the derivative Schrodinger equation also is discussed its relation to
the Ishimori-II model. Some pecular soliton solutions of nonlinear Schrodinger
type equations are given and discussed.
\end {abstract}
\vspace{0.5cm}

Four years ago a paper on exact solutions of a time-dependent
Schr\"odinger equation with self consistent potentials was published in
the journal, Particles and Nuclei [1]. In this work the method is
developed to construct such solutions along with the nonlinear
equations they obey.

Though the paper was of a survey character it contained a series of
original results. Talking at International Conferences and in some European
Centres
I realized that these results were practically unknown to the
audience. This makes me to look over the paper [1] and try to rewrite
its main body adding some recent development in this direction. So the
first part of the present paper reproduces in a more simple way (up to
author's opinion) the results of [1] related to studying the
time-dependent linear Schor\"odinger Equation (TLSE)
\begin{equation}
[i\partial_t-\partial^2_x+U(x,t)]\Psi (x,t,k)=0.
\end{equation}
Only solution-type solutions will be discussed for the sake of
simplicity. For finite zone solutions see [2,3]. And in the second part
we give an outlook of how to extend this approach to include the
equation (Derivate Schr\"odinger Equation, TDSE)
\begin{equation}
[i\partial_t-\partial^2_x+iU(x,t)\partial_x]\Psi (x,t,k)=0
\end{equation}
which gets popular now due to its connection to some 2+1 dimensional
models (for Ishimori-II and modified KP models, see, eg.[2]).

First of all we stress that there are two different levels of
consideration: linear and nonlinear. On the first we shall find a
special class of localized reflectionless (Bargman's) potentials along
with their wave-functions. They are defined, conditionally speaking, by
certain "spectral data" (SD), namely, by a set of complex numbers
\ $\kappa_i\ \ i=1,N$\ and a complex valued \ $N\times N$\ matrix \
$c_{ij}$\ (normalization matrix). We give the conditions which \
$\kappa_i$\ and \ $c_{ij}$\ have to satisfy in order to the potential \
$U$\ to be real and nonsingular together with w.f.\ $\Psi$. We also
discuss degeneration of the solution with respect to the S.D. and give
two possible representations of the w.f.viz.,the polynomial one and the
rational (pole-type) one. Asymptotic properties of the solutions allow
us to judge of the structural units (bricks) which the solutions are
constituted of.

On the second nonlinear level self-consistency conditions are found
which relate the potential to the w.f. and its residues. Here boundary
conditions for nonlinear fields play the crucial role and the fields
are given as direct sums of the structural units.

Some simple concrete examples are discussed with special emphasis on
peculiar bi-pole solutions.
\section{Linear level.}
1. We look for the soliton like solutions of eq.(1) via the plane wave
Ansatz
\begin{equation}
\Psi(x,t,k)=P(x,t,k)e^{ik(x+kt)}\equiv
(k^n+a_{n-1}(x,t)k^{n-1}+\cdots+a_0(x,t))e^{ik(x+kt)}
\end{equation}
and by specifying the SD: poles \ $\kappa_i,\ \ i=\overline{1,N}$\ with nonzero
imaginary parts and a \ $N\times N$\ matrix \ $\alpha_{ij}$.

Generalizing the notion of the Beiker-Akhiezer function we assume the
w.f.\ $\Phi$\ to satisfy the following \ $N$\ conditions
\begin{equation}
\Psi(\bar\kappa_i)=-\sum_{j=1}^N b_{ij}\Psi(\kappa_j),\quad i,j=1,\cdots,N
\end{equation}
\underline{\large\bf First proposition.}
\vspace{0.5cm}

Given with (3) and (4) w.f. \
$\Psi(k,x,t)$\ defines the potential \ $U(x,t)$\ as follows
\begin{equation}
U(x,t)=2i\partial_xa_{N-1}(x,t)
\end{equation}
The proof is very easy, and can be found in [1] (and even earlier in [3]
for the B-A. function).

The representation (3) of the w.f. we call polynomial and now it's
of use to introduce the other
\begin{equation}
\Psi=\frac{\Psi(k,x,t)}{\prod\nolimits_{i=1}^N(k-\kappa_i)}\equiv
\left\{1+\sum^{N}_{j=1}\frac{r_j(x,t)}{k-\kappa_j}\right\}e^{ik(x+kt)}
\end{equation}
which naturally may be named the rational or pole-type representation.
We need both of them in what follows.

In the pole type representation eq.(4) assumes the form
\begin{equation}
\Psi(x,t,\bar\kappa_i)=-\sum_{j=1}^N c_{ij}\Psi_j(x,t)
\end{equation}
with
\begin{equation}
\Psi_j=res_{k=\kappa_j}
\Psi(x,t,k)=\lim_{k\to\kappa_j}[(k-\kappa_j)\Psi(x,t,k)]
\end{equation}

and
\begin{eqnarray}
c_{ij}&=&b_{ij}\frac{R'(\kappa_j)}{R(\bar\kappa_j)}\nonumber\\
&&\\
R(k)&=&\prod_{j=1}^N(k-\kappa_j),\quad R'=\frac{d}{dk}R(k)
\end{eqnarray}

The SD \ $\kappa_i$\ and \ $c_{ij}$\ are still \underline{arbitrary}
 (with to the above exception). \\
\vspace{0.5cm}
\underline{\large\bf Second proposition}. \\
In order to the potential \ $U(x,t)$\ be real and non-singular for all
real  $(x,t)$ the following conditions are sufficient

1)The matrix \ $c_{ij}$\ in (8) should be anti-Hermitian: \ $c=-c^{+}$.

2)Let the poles \ $\kappa_i$\ be arranged such that \ $Im\ \kappa_i>0$\
for \ $i=\overline{1\cdot p}$ and \ $Im\ \kappa_j<0$\ for \
$j=\overline{p+1,N}$\ then the Hermitian matrix \ $(i^{-1}c_{ij}),\ \
i,j=\overline{1,p}$\ should be positive definite and the \ $((i^{-1}c_{ij}),
\ \  i,j=\overline{p+1,N}$\ negative definite. \\
\vspace{0.5cm}
The proof is more sophisticated and also given in [1].

These conditions gives the first real limitation on the location of the
poles related to the form of \ $c_{ij}$.

The following properties of the solution obtained have to be mentioned.

a) The solution is degenerate. In polynomial
representation (3) both the w.f.\ $\Psi(x,t,k)$\ and the
potential \ $U(x,t)$\ are \ $2^N$-fold degenerate with respect to
changing the SD.

In the pole-type representation only the potential \ $U(x,t)$\ is left \
$2^N$-fold degenerate, i.e.  for \ $2^N$\ definite but different sets
of the SD we have the same potential \ $U(x,t)$.

Transformations from one set to another goes as
follows [1].
Let the matrix\ $\alpha_{ij}$\ be given in the block form
\begin{equation}
(b_{ij})=\left(\begin{array}{ll}
\alpha_{+}&\beta\\ \gamma&\alpha_-\end{array}\right)
\end{equation}

where the square matrices \ $\alpha_+$\ and\ $\alpha_-$\
are \ $p\times p$\ and\ $(N-p)\times (N-p)$\ dimensional
respectively (recall that \ $Im~\kappa_i>0$\ for \
 $i=\overline{1,p}$, and \ $Im~\kappa_i<0$\ if \
 $i=\overline{p+1,N}$) and \ $det~\alpha_-\not=0$, then the
transformations are \ $\{\kappa_i,b_{ij}\}\Longrightarrow
\{\kappa_i',b'_{ij}\}$\ such that
\begin{equation}
\kappa_i'=\left\{\begin{array}{l}
\kappa_i \ \ \mbox{for i=}\ \ \overline{1,p}\\\bar\kappa_i\ \ \mbox{for i=}\ \
\overline{p+1,N}
\end{array}\right.
\end{equation}
and
\begin{equation}
(b'_{ij})=\left(\begin{array}{cc}
\alpha_+-\beta\alpha^{-1}_-\gamma&-\beta\alpha_-^{-1}\\
\alpha_-^{-1}\gamma&\alpha_-^{-1} \end{array}\right)
\end{equation}

b)One can consider the asymptotic behavior of the solution in \ $x$\ and
\ $t_i$\ at arbitrary \ $N$.

i) In the simplest case \ $N=1,~ \kappa =\alpha +i\beta$\ \ the potential
assumes
soliton like form
\begin{equation}
U=-2\beta^2cosh^{-2}\beta(x-x_0+2\alpha t),
\end{equation}
and w.f. is
\begin{eqnarray}
\Psi&=&\left[1+\frac{i\beta}{k-\kappa}(1+th\beta(x-x_0+2\alpha
t))\right]e^{ik(x+kt)}\nonumber\\
&&\\
2i\beta c&=&-e^{2\beta x_0}\nonumber
\end{eqnarray}
For \ $N>1$, all \ $Im\ \kappa_i>0$\ and \ $Re\ \kappa_i\not=Re\ \kappa_j$ \
if~$i\not=j$\ the potential asymptotically decays in a direct sum of the
potentials (14). Hence such \ $N=1$\ potentials can be regarded as
simple construction bricks for complexes with \ $N>1$. Below we call
them the {\it \bf solibricks}.

ii) The next fundamental construction units are {\it \bf strings}:\\
 $N>1$\, \ $Re\ \kappa_i=Re\ \kappa_j$\ which are periodic or quasiperiodic in
time
solutions [4] sometimes named the {\it breathers}. Both types of solutions are
well-known.

iii) Now we proceed to a {\bf new} type of construction units the {\it \bf
bions}. They
are defined by an of diagonal matrix c, e.g. in the \ $N=2$\ case
\begin{equation}
c=\left(\begin{array}{rl}0&1\\-1&0\end{array}\right)
\end{equation}
and are given by [5]:

\begin{equation}
\Psi =\left(1+ \frac{c_3cos(q\xi +\Omega x+\theta_3)+c_4e^{\beta\xi}}
{c_2cosh(\beta\xi +\theta_2)+c_1cos(q\xi +\Omega
x+\theta_1)}\right)e^{ik(\xi+k'x)}
\end{equation}
with
\begin{eqnarray*}
c_1&=&-\left|\frac{c}{\tilde\kappa_{12}}\right|,\qquad
e^{\theta_2}=\left|\frac{\tilde\kappa_{12}}{c\kappa_{12}}\right|\sqrt{\frac{1}{\tilde\kappa_{11}\tilde\kappa_{22}}}\\
c_2&=&\left|\frac{c\kappa_{12}}{\tilde\kappa_{12}}\right|\sqrt{\frac{1}{\tilde\kappa_{11}\tilde\kappa_{22}}},\qquad e^{2i\theta_1}=-\frac{c\kappa_{12}}{\bar c\bar \kappa_{12}}\\
c_3&=&- |c|\sqrt{\frac{1}{(k-\kappa_1)(k-\kappa_2)}},\qquad
e^{2i\theta_3}=-\frac{c(k-\kappa_2)}{\bar c(k-\kappa_1)}\\
c_4&=&-\frac{1}{2}\left(\frac{\bar\kappa_{21}}{(k-\kappa_1)\tilde\kappa_{12}\tilde\kappa_{22}}+\frac{\bar\kappa_{12}}{(k-\kappa_2)\tilde\kappa_{21}\tilde\kappa_{11}}\right)^{\frac{1}{2}}
\end{eqnarray*}

The parameters defining solution (17) are
$$\kappa_j =\alpha_j + i\beta_j $$
$$ q=\alpha_2 -\alpha_1 , \qquad \Omega =\omega -qv, \qquad k'=k-v, \qquad
\beta =\beta_1 +\beta_2 ,$$
$$\omega =\alpha_2^2 -\alpha_1^2 +\beta_1^2 -\beta_2^1 ,\qquad \xi =t+vx,$$
$$v=2\frac{\alpha_1\beta_1 +\alpha_2\beta_2}{\beta_1 +\beta_2} $$
$$\tilde\kappa_{ij} =\bar \kappa_i -\kappa_j , \qquad \kappa_{ij} =\kappa_i
-\kappa_j$$
and $\alpha_i$ give velocities of the constituents, $\beta_i$ their ``masses''.

These solutions also are periodic or quasiperiodic, and we call them the
bions. The nature of breathers and bions is strongly different,
especially from the physical point of view (it will be evident on the
nonlinear level). Though a breather and a bion are formed with two
solibricks and both quasiperiodic in time the first can be easily
destroyed to produce the constituents, for the second such a process is
strongly forbidden. The constituents of the bion behave like quarks in a
meson. Therefore we found so far three types of structural units:
solibricks, breathers and bions (next are possible), and the potential
asymptotically assumes the form
\begin{equation}
U(x,t)_{t\rightarrow\infty}=\sum solibricks+\sum breathers+\sum bions+\cdots
\end{equation}
These results are directly applicable to construct soliton-like
solutions of the Davey-Stewartson-I equation. First attempts to find such
solutions were done in [6] and [7] and the $dromions$ were born. In [8]
some new results based on the approach proposed were obtained. In this
sense one can consider the above technique as a constructive way to
produce solutions in the scope of 2+1 dimensional KP and DS-I models.

2. Consider now the equation (2)
\begin{equation}
(i\partial_t-\partial^2_x\pm iU\partial_x)\Psi(x,t,k)\equiv L\Psi(x,t,k)=0
\end{equation}
Adopting the above procedure to (2) we construct its
solutions through the Ansatz slightly different from (3)
\begin{equation}
\Psi(x,t,k)=Q_Nexp[ik(x+kt)]
\end{equation}
with
\begin{equation}
Q_N=a_Nk^N+\cdots+a_1k+1
\end{equation}
Specify again \ $N$\ poles \ $\kappa_i$\ and complex \
$N\times N$\ constant matrix \ $b_{ij}$\ we can prove the
first proposition [9] for the w.f. \ $\Psi$\ (18)
satisfying (4) but now
\begin{equation}
U(x,t)=2i\partial_xlna_N
\end{equation}
One can introduce again the pole-type representation
\begin{equation}
\hat\Psi=a_N\left\{1+\sum_{j=1}^N\frac{r_j(x,t)}{k-\kappa_j}\right\}e^{ik(x+kt)}
\end{equation}
to obtain exactly the equations (7) for the function
\begin{equation}
\Psi=a_N^{-1}\hat\Psi
\end{equation}
and then
\begin{equation}
a_N=\frac{(-1)^N}{\prod\nolimits_{j=1}^N\kappa_j}\ \
\frac{1}{1-\sum\nolimits_{j=1}^N\frac{r_j}{\kappa_J}}
\end{equation}
Denoting \
$\Psi_0=\Psi(x,t,k=0)=1-\sum_{j=1}^N\frac{r_j}{\kappa_j}$\ we
have
\begin{equation}
a_N=\prod_{j=1}^N\left(\frac{-1}{\kappa_j}\right)\ \ \frac{1}{\Psi_0}
\end{equation}
and
\begin{eqnarray}
U&=&-2i\partial_xln\Psi_0\\
\hat\Psi&=&\prod_{j=1}^N\left(\frac{-1}{\kappa_j}\right)\ \
\frac{1}{\Psi_0}\Psi(x,t,k)
\end{eqnarray}
 From (25) it follows that \ $U=\bar U$\ when
\begin{equation}
\left|1-\sum_{j=1}^N\frac{r_j}{\kappa_j}\right|=const
\end{equation}
General conditions are still unknown which the SD should satisfy
in order to obtain a real and non-singular potential \
$U(x,t)$.

So we give a simplest \ $N=1$\ example and find the
connection of the such obtained one soliton solution with
that of the Ishimori-II equation.

In the one soliton case system (4) gives
\begin{equation}
a=(-\frac{1}{\kappa})\frac{1+be^{i(\theta
-\bar\theta)}}{1+b\frac{\kappa}{\bar\kappa}e^{i(\theta -\bar\theta)}}
\end{equation}
The potential \ $U$\ is real when \ $|a|=const$\ or
\begin{equation}
b\kappa =\overline{b\kappa}
\end{equation}
or \ $(b=b_1+ib_2)$
\begin{equation}
b_1\beta +b_2\alpha =0
\end{equation}

By differentiating \ $lna$\ (30) we arrive at
\begin{eqnarray}
V&=&\frac{8|\alpha|\beta^2sgn\alpha}{4\alpha^2cosh^2\eta
+\beta^2e^{-2\eta}},\quad b_1=e^{-2\beta x_0}>0\\[0.3cm]
U&=&-\ \ \frac{8|\alpha|\beta^2sgn\beta}{4\alpha^2sinh^2\eta
+\beta^2e^{-2\eta}},\quad b_1=-e^{-2\beta x_0}<0\\[0.3cm]
\eta&=&\beta(x+2\alpha t+x_0)\\[0.3cm]
\mbox{also we have}&&\nonumber\\[0.3cm]
\Psi&=&\left(1-\frac{k}{\bar\kappa}\frac{1+be^{-2(\eta - \beta x_0)}}{1+\bar
be^{-2(\eta -\beta x_0)}}\right)e^{ik(x+kt)}\\[0.3cm]
\mbox{If one sets}\ k&=&\bar\kappa\ \ then\nonumber\\[0.3cm]
\Psi&=&2i\frac{\sqrt{b_1}\beta e^{i\theta}}{2\alpha cosh\eta+i\beta
e^{-\eta}},\quad \theta=i\alpha x+i(\alpha^2-\beta^2)t
\end{eqnarray}
Fixing \ $k=\kappa$\ we come to the same expression up to the constant factor.

It should be pointed out that a general form for the w.f. of time dependent
eq.(2) is that of (36) with arbitrary complex number k, so the solution is a
five real parameter \ $(\alpha,\beta,b_1,k_1,k_2)$\ function.

3. We now utilize the solutions found in order to obtain such for the
Ishimori-II model
\begin{eqnarray}
\vec S_t(x,y,t)+\vec S \wedge(\vec S_{xx}+\vec S_{yy})+\varphi_x\vec
S_y+\varphi_y\vec S_x=0\\
\varphi_{xx}-\varphi_{yy}+2\vec S(\vec S_x\wedge\vec S_y)=0
\end{eqnarray}
where \ $\vec S=(S_x,S_y,S_z),\ \ \vec S^2=1$\ \ and \ $\varphi(x,t)$\ is a
real function.

To treat eqs.(38), (39) we shall use the results of work [2] where solutions of
the problem was reduced to solutions of two linear equations of type (2),
namely
\begin{eqnarray}
iX_t(\xi,t)+\frac{1}{2}X_{\xi\xi}+iU_2(\xi,t)X_{\xi}=0\\
iY_t(\eta,t)+\frac{1}{2}Y_{\eta\eta}-iU_1(\eta,t)Y_{\eta}=0
\end{eqnarray}
with \ $\xi=\frac{1}{2}(x+y),\ \  \eta=\frac{1}{2}(y-x)$\ and real potentials:
\ $U_i=\overline U_i.$

The special class of solutions of (36) related to degenerate spectral data
(factorized) are given by the formulas [2]: (N=1)
\begin{eqnarray}
S_x+iS_y&=&2\frac{XY}{|1-AB|^2}(1+\bar AB),\quad S_-=\bar S_+\\[0.3cm]
S_3&=&-\left(1+2\frac{(A+\bar A)(B+\bar B)}{|1-AB|^2}\right),\\[0.3cm]
\varphi(\xi,\eta,t)&=&2iln(det\Delta)+2\partial_{\xi}^{-1}U_2(\xi,t)+2\partial_{\eta}^{-1}U_1(\eta,t)\\[0.3cm]
A&=&\int_{-\infty}^{\eta}d_y\bar Y(y,t)\partial_yY(y,t)\\[0.3cm]
B&=&-\int_{-\infty}^{\xi}d_xX(x,y)\partial_x\bar X(x,t)\\[0.3cm]
\Delta&=&\frac{1-\bar{AB}}{1+AB}
\end{eqnarray}
 one can easily see from (40), (41) that
$$Y(x,t)=\bar X(x,-t)$$
Consider the case \ $b_1>0, \ k=\kappa$.\ Then
\begin{eqnarray}
X&=&\frac{e^{i\alpha_1x+i(\beta_1^2-\alpha_1^2)t}}{2\alpha_1coshz_1+i\beta_1e^{-z_1}},\quad z_1=\beta_1(x-2\alpha_1t+x_0)\\[0.3cm]
Y=\bar
X(y,-t)&=&\frac{e^{-i\alpha_2y+i(\beta_2^2-\alpha_2^2)t}}{2\alpha_2coshz_2-i\beta_2e^{-z_2}},\quad z_2=\beta_2(y+2\alpha_2t+y_0)\\[0.3cm]
A&=&\frac{1}{2}\ \
\frac{1-i\frac{\alpha_2}{\beta_2}(1+e^{2z_2})}{4\alpha_2^2cosh^2z_2+\beta_2^2e^{-2z_2}}\\[0.3cm]
B&=&-\frac{1}{2}\ \
\frac{1-i\frac{\alpha_1}{\beta_1}(1+e^{2z_1})}{4\alpha_1^2cosh^2z_1+\beta_1^2e^{-2z_1}}
\end{eqnarray}
The solution (42)--(51) is a one soliton solution which moves with the velocity
\ $\vec v=(2\alpha_1,-2\alpha_2)$.\ One can proceed to the moving coordinate
frame to obtain the solution at rest.
\section{Nonlinear level}

Here we give a short outlook of the results in the scope of (1). The main
problem is to find a relation connecting the potential \ $U(x,t)$\ and the w.f.
\ $\Psi(x,t,k)$\ or its residues \ $\Psi_i(x,t)$. Since by construction we deal
with meromorphic functions it is naturally of use to apply the residue
theorem and calculate residues sum.

This prompts the form of a rational function \ $E(k)$\ using which one can
find the self consistency condition via calculating the residues of the
function
\begin{equation}
\Omega=E(k)\overline{\Psi(x,t,\bar k)}\Psi(x,t,k)
\end{equation}
If we specify \ $E$\ as the polynomials:
\begin{eqnarray}
1)\ E_1&=&k\\
2)\ E_2&=&k^2+ak\\
3)\ E_3&=&k_3+2bk^2+2dk
\end{eqnarray}
we arrive at the following relations respectively
\begin{eqnarray}
1)\ {U}&=&-2F(x,t)\\
2)\ U_t+aU_x&=&-2\partial_xF(x,t)\\
3)\
(\partial_t^2-\frac{1}{3}\partial^4_x)U&+&2(U^2)_{xx}+\frac{8}{3}bU_{xt}+\frac{8}{3}dU_{xx}=-\frac{8}{3}\partial_x^2F(x,t)
\end{eqnarray}
where \ $F(x,t)$\ is the quadratic form
\begin{equation}
F(x,t)=\sum^N_{ij}\bar\Psi_iE_{ij}\Psi_j
\end{equation}
in which
\begin{equation}
E_{ij}=(E(\bar\kappa_j)-E(\kappa_j))c_{ij}
\end{equation}
is the Hermitian matrix.

This matrix along with the set of the poles \ $\kappa_i$\
completely define solutions of the nonlinear equation
\begin{equation}
(i\partial_t-\partial^2_x+U(x,t))\Psi_i(x,t)=0
\end{equation}
with corresponding self consistency relations (56), (57) or (58).

In eq.(61) nonlinear fields are just the residues of the w.f.
\ $\Psi(x,t,k)$\ and they possess ``right" asymptotic behaviors at
\ $x\longrightarrow\pm\infty$\ at certain sets of the SD (see proposition 2).
In
these cases
\begin{equation}
\Psi_i(x\longrightarrow\pm\infty)\longrightarrow 0
\end{equation}
and we have the  nonlinear problem with \underline{Trivial Boundary
Conditions} (TBC). For other sets of the SD \ $\Psi_i$\ infinitely grow and
usually are
not interesting from the physical point of view (at least for homogeneous
systems).

In order to extend our treatment to the case of nontrivial (so-called
Condensate Boundary Conditions)(CBC)
\begin{equation}
\left|\Phi_i(x,t)\right|\longrightarrow const
\end{equation}
one has to consider instead of (53)--(55) the following function \
$E=\tilde E$\ with
\begin{equation}
\tilde E=\sum_{i=1}^n\frac{\varepsilon_ib_i^2}{k-k_i}+E_{\imath}\quad
 \imath=1,2,3
\end{equation}
and \ $\varepsilon_i=\pm 1, \ b_i\ and \ k_i$\ are arbitrary real
numbers.

Calculating the residues of \ $\Omega$\ (see (52)) we found
equations (56)--(58) but \ $F(x,t)$\ is now
\begin{equation}
\tilde
F(x,t)=\sum_{ij}^N\bar\Psi_iE_{ij}\Psi_j+\sum_{m=1}^n\varepsilon_m(|\Phi_m|^2-b_m^2)
\end{equation}
and nonlinear fields
\begin{equation}
\Phi_i(x,t)=b_i\Psi(x,t,k=k_i)
\end{equation}
are the w.f. at fixed points \ $k=k_i$.

 From asymptotic behavior at  $x\rightarrow\pm\infty$\ it
follows (see [1]) that  $\Psi(t,k_i x\rightarrow\pm\infty)=1$\
and as a result we have CBC (63) for the nonlinear fields \
$\Phi_i(x,t)$.

In general case one can consider a \ $(n+m)$\ component vector
field
\begin{equation}
\varphi=\left(\begin{array}{c}
\Psi_1\\
\vdots\\
\Psi_n\\
\Phi_1\\
\vdots\\\Phi_m
\end{array}\right)
\end{equation}
satisfying the equation
\begin{equation}
\left[i\partial_t-\partial_x^2+U(x,t)\right]\varphi=0
\end{equation}
with self consistency conditions (56), (57) or (58) and (65).

It is clear that in the case of pure condensate fields the form \
$F(x,t)=\sum_{ij}^N\bar \Psi_iE_{ij}\Psi_j$\ must vanish for every nonzero
solibricks \ $\Psi_i$.\ This puts extra (nonlinear) restrictions on the SD,
namely, on the location of the poles. Consider for example a well--known case
of
the scalar NSE
\begin{equation}
(i\partial_t-\partial_x^2+\varepsilon\ (|\Phi|^2-b^2)\Phi=0
\end{equation}
with \ $N=1$\ and CBC (63).

In this case we have
\begin{equation}
\tilde E(\bar\kappa_1)-\tilde E(\kappa_1)=0
\end{equation}
or
\begin{equation}
(\bar\kappa_1-\kappa_1)(\varepsilon\ \frac{b^2}{|\kappa_1-k_1|^2}-1)=0
\end{equation}

One can easily see from (71) that the equation
\begin{equation}
\varepsilon\ \frac{b^2}{|\kappa_1-k_1|^2}=1
\end{equation}
has a solution when \ $\varepsilon=1$,\ i.e. only in the case of the repulsive
NSE,
moreover poles allowed are on the circle \ $|\kappa_1-k_1|^2=1/b^2$.\
In framework of the attractive NSE\ $(\varepsilon=-1)$\ a
one--pole condensate solution is absent.

As the second example, we consider a new two--pole solution
namely bions (17). Then instead of (71) we have
\begin{equation}
\tilde
E(\bar\kappa_1)-E(\kappa_2)=0\quad\mbox{or}\quad(\bar\kappa_1-\kappa_2)\left(\varepsilon\ \frac{b^2}{(\bar\kappa_1-k_1)(\kappa_2-k_1)}-1\right)=0
\end{equation}

The first solution \ $\bar\kappa_1=\kappa_2$\ is just the
well--known Zakharov--Shabat breather (string solution), so we
discuss solutions of the equation
\begin{equation}
\kappa_2=k_1+\varepsilon\ \frac{b^2}{|\kappa_1-k_1|^2}(\kappa_1-k_1)
\end{equation}
 From (74) it follows that the condition \ $Im\ \kappa_2\cdot Im\ \kappa_1<0$\
needed for nonsingularity of the solution
(Proposition 2) is valid when \ $\varepsilon\ <0$,\ i.e. in the scope of the
attractive NSE.

So the qualitative conclusion is as follows:

In the system with condensate (plane--wave boundary conditions) one--pole
solutions (kink) exist for the repulsive type interaction and absent otherwise
, meanwhile two--pole solutions (bions) appear in the scope of the attractive
NSE as the elementary nonlinear excitation and consist of the
\underline{unvisible} (like quarks) constituents, the solibricks.

Finaly we note that in the scope of the above technique vector nonlinear
Schr\"odin-   \
ger equations with nondiagonal  potentials [9] can be treated, too.
For example, the following simplest system
\begin{equation}
i\phi_{it}-\phi_{ixx}+U(x,t)\phi_i =0  \ \ \ i=1,2
\end{equation}
$$U(x,t) =  \bar \phi_1 \phi_2 +\phi_1 \bar \phi_2 $$
has the solution
\begin{equation}
\phi_i =\frac{A_i e^{i(q_1x+\omega_1t)}ch(\beta_1(x+v_1t)+b_i) +
B_i e^{i(q_2x+\omega_2t)}ch(\beta_2(x+v_2t)+a_i)}{B_1ch(\beta^+(x+v^+t)+h_1)
+B_2ch(\beta^-(x+v^-t)+h_2)+B_3cos(qx+\omega t+\omega_0)}
\end{equation}
with the trivial boundary conditions
\begin{displaymath}
{\phi_1 \choose \phi_2}_{x \to \mp \infty}= { 0 \choose 0}
\end{displaymath}
and the coefficients similar to those defined in (17).

This paper is dedicated to the kind memory of Professor M.K.Polivanov talks
with whom always were interesting and often instructive.
\vspace{1.5cm}

{\large\bf References}
\vspace{1cm}

\begin{enumerate}
\item B.Dubrovin, I. Krichever, T. Malanynk, and V.Makhankov. Exact
Solutions of Time--Dependent Sch\"orodinger Eq. with Self--Consistent
Potential. {\it Sov. J. Part. Nuclei,} {\bf 19 (3) (1988) 252-269}.
\item B.Konopelchenko and V.Dubrovsky. Coherent Structures for the Ishimori
equation. Localized Solutions with Stationary Boundaries I. {\it Physica} {\bf
D48 (1991) 367-395};  Time-Dependent Boundaries II. {\it ibid} {\bf D55 (1992)
1-13}.
\item I.Krichever. Spectral Theory of Finite Zone Nonstationary Schr\"odinger
Operators. Nonstationary Pierls Model.{\it Funct. Anal. Applic.} {\bf 20 (1986)
42-54.}
\item V.Zakharov, A.Shabat. Exact theory of two-dimensional self-focusing and
one-dimensional automodulation of waves in nonlinear media. {\it ZhETF} {\bf 61
(1971) 118-134.}
\item V.Makhankov and S.Slavov. In {\it Solitons and Applications}, eds.
V.Makhankov, V.Fedyanin and O.Pashaev World Scientific, Singapore 1990,
pp.107-113.
\item M.Boiti, J.Leon, L.Martina and F.Pempinelli.{\it Phys. Lett.} {\bf A132
(1988) 432.}
\item A.Fokas, and P.M.Santini. {\it Phys. Rev. Lett.} {\bf 63 (1989) 1329.}
\item M.Boiti,L.Martina, O.Pashaev and F.Pempinelli. Dynamics of
multidimensional solution. {\it Phys. Lett,} {\bf A160 (1991) 55.}
P.M.Santini. {\it Physica}, {\bf D41 (1990) 26.}
\item V.Makhankov. On Soliton Type Solutions of one Linear Time-Dependent
Equation and the Ishimori-II Model with Time-Dependent Boundary Conditions.
{\it JINR,} {\bf E4-92-208}, Dubna, 1992.
\item A.Fordy, and P.Kulish. {\it Comm. Math. Phys.} {\bf 89 (1983) 427.}
\end{enumerate}
\end{document}